\documentclass[aps,pre,showpacs,showkeys,twocolumn,groupedaddress]{revtex4-1}

\begin{document}


\title{Questing for an optimal, universal viral agent for oncolytic virotherapy}


\author{L R Paiva} 
\email{leticia.paiva@ufv.br}
\author{M L Martins}
\email{mmartins@ufv.br}
\altaffiliation{National Institute of Science and Technology for Complex Systems, Brazil}
\author{S C Ferreira Jr}
\altaffiliation{On leave at Departament de F\'{\i}sica y Enginyeria Nuclear, Universitat Polit\'ecnica de Catalunya, Spain}
\affiliation{Departamento de F\'{\i}sica, Universidade Federal de Vi\c{c}osa, 36570-000, Vi\c{c}osa, MG, Brazil}


\date{\today}

\begin{abstract}

One of the most promising strategies to treat cancer is attacking it with viruses designed to exploit specific altered pathways. Here, the effects of oncolytic virotherapy on tumors having compact, papillary and disconnected morphologies are investigated through computer simulations of a multiscale model coupling macroscopic reaction diffusion equations for the nutrients with microscopic stochastic rules for the actions of individual cells and viruses. The interaction among viruses and tumor cells involves cell infection, intracellular virus replication and release of new viruses in the tissue after cell lysis. The evolution in time of both viral load and cancer cell population, as well as the probabilities for tumor eradication were evaluated for a range of multiplicities of infection, viral entries and burst sizes. It was found that in immunosuppressed hosts, the antitumor efficacy of a virus is primarily determined by its entry efficiency, its replicative capacity within the tumor, and its ability to spread over the tissue. However, the optimal traits for oncolytic viruses depends critically on the tumor growth dynamics and do not necessarily include rapid replication, cytolysis and spreading currently assumed as necessary conditions to a successful therapeutic outcome. Our findings have potential implications on the design of new vectors for the viral therapy of cancer.
\end{abstract}

\maketitle
\section{INTRODUCTION}
Despite the progress that has been made in imaging, diagnosis, treatment and understanding of cancer, the survival rates of patients with metastatic or recurrent neoplasias as well as patients with tumors at unresectable locations are still poor, and new therapeutic strategies are needed \cite{Ketola}. Furthermore, occult dormant micrometastasis raise new challenges concerning their reactivation and evolution into clinically manifested disease, particularly after primary tumor resection \cite{Naumov}.

Oncolytic virotherapy is claimed to be a promising anti-cancer strategy, because it can provide a locoregional control or even eradication of tumors without cross-resistance with standard therapies. Oncolytic viruses are able to selectively infect and kill tumor cells by exploiting the same cellular defects that promote tumor growth \cite{Liu}. To date, several different viruses are known to be selectively oncolytic \cite{Harrington}. An archetype is the adenovirus commonly used in gene therapy and oncolytic therapy experiments \cite{Yamamoto,Relph}. Some studies indicate that the virus effectiveness strongly depends on the specific cancer cell line \cite{Ketola}. So, one can hypothesize that for each tumor there must exist an oncolytic virus that maximizes the therapeutic success.

Several mathematical models have been proposed to study virotherapy \cite{Friedman,Paiva,Wodarz,Reis}. They emerge as valuable tools to provide quantitative understanding of the major mechanisms that affect anti-cancer treatment based on viruses and to select parameter ranges that enhance its therapeutic success.  Specially because, in face of the nonlinearities and complexity involved in cancer progression and its interaction with oncolytic viruses, intuitive reasoning alone may be insufficient. The current paradigm is that the viral life cycle should lead to rapid replication, cytolysis and spread \cite{Parato}. Nonetheless, the limitations and validity of this appealing, intuitive, and seemingly logical paradigm remains unclear. Indeed, at least one mathematical result \cite{Paiva} indicates that viruses able to destroy tumor cells very fast does not necessarily lead to a more effective control of tumor growth. More precisely, it was shown that a successful virotherapy of compact tumors requires both highly spreading viruses and an optimal range of viral cytotoxity, i. e, neither a too short nor a very large time for the lysis of cancer cells. However, this result was derived assuming that the viruses are a continuous field whose dynamics is described by a reaction-diffusion equation. Furthermore, the changes in the stationary local concentrations of viruses due to their entry at cell infections were neglected and the viral burst size used, equal to the initial viral load, was orders of magnitude greater than the larger ones observed for real oncolytic viruses.

In the present paper, a multiscale model based on the approach proposed by Ferreira et al. \cite{FerreiraPhysA} was modified to take explicitly into account the individual, discrete nature of the oncolytic viruses and applied to evaluate the efficacy of virotherapy against compact, papillary and diffuse (disconnected) solid tumors. Considering virus as discrete agents provides a more realistic description of virus entry and replication, key processes involved in oncolytic virotherapy. Our main goal is provide useful insights about how to match the oncolytic virus and the tumor type in order to get the better outcome for the therapy.

\section{MODEL}
Figure $1$ illustrates the multiple agents and processes involved in the model at distinct time and length scales. The tissue is modelled by a square lattice fed through a single capillary vessel at its top. Four different cell types (normal and dead cells, uninfected and infected tumor cells) and 
an oncolytic virus are considered. These individual agents are described by their populations $\sigma_n$, $\sigma_d$, $\sigma_c$, and $\sigma_v$,  respectively, at every site $\mathbf{x}$. In contrast to the normal and dead cells, one or more uninfected or infected cancer cells can pile up in a
 given site, reflecting the fact that the division of tumor cells is not constrained by contact inhibition. In turn, since the viruses are very small
particles in comparison with cells, there is no constraint on their population.

The nutrients, diffusing from the capillary vessel throughout the tissue, are divided into two groups: those that limits cell replication but are 
not demanded for cell survival ($j=1$) and those essential to maintain the basic cell functions and whose deprivation can induce death ($j=2$). 
Both nutrient types are described by continuous fields $\phi_j(\vec{x},t)$, which evolve in space and time according the simplest (linear with 
constant coefficients) dimensionless reaction-diffusion equations

\begin{equation}
\frac{\partial \phi_j}{\partial t}= \nabla^2 \phi_j - \alpha^2 \phi_j \sigma_n - \lambda_j \alpha^2 \phi_j \sigma_c.
\label{dif_nut}
\end{equation}
The factors $\lambda_j$ take into account distinct nutrient uptake rates for normal and cancer cells. The parameter $\alpha$ sets up a characteristic length scale for nutrient diffusion in the normal tissue (see reference \cite{Ferreira2002} for the complete variable transformations leading to this dimensionless equation). Eq. (\ref{dif_nut}) obeys a periodic boundary condition along the direction parallel to the capillary and a Neumann boundary condition at the border of the tissue. At the capillary vessel the nutrient concentrations are $\phi_j=1$ (continuous and fixed supply).

Each uninfected cancer cell, randomly selected with equal probability, can carry out one of four actions: 

\noindent 1-\textbf{Mitotic replication}, with a probability

\begin{equation}
P_{div}=1-\exp{\left[ -\left( \frac{\phi_1}{\theta_{div} \sigma_c} \right)^2 \right]},
\label{pdiv}
\end{equation}
an increasing function of the concentration per cancer cell of the nutrients $\phi_1$. The daughter cell randomly occupies one of their normal or necrotic nearest neighbor sites, whether there exists any, or piles up at its mother site. In the simulations, we observed at most three or four cancer cells at the same site simultaneously.

\noindent 2-\textbf{Death}, with a probability

\begin{equation}
P_{del}=\exp{\left[ -\left( \frac{\phi_2}{\theta_{del} \sigma_c} \right)^2 \right]},
\label{pdel}
\end{equation}
that increases with the scarcity of nutrients $\phi_2$ essential to sustain the cell metabolism.

\noindent 3-\textbf{Migration}, with a probability

\begin{equation}
P_{mov}=1-\exp{\left[ - \sigma_c \left(\frac{\phi_2}{\theta_{mov}} \right)^2 \right]},
\label{pmov}
\end{equation}
that increases with the local population of cancer cells and the nutrient concentration per cell. A probability increasing with the nutrient concentration is justified by the necessity of nutrients for cell motility and, in addition, by the degradation of the extracellular matrix nearby the tumor surface that releases  several chemicals which promote cell migration and proliferation. This hypothesis is consistent with experimental data in multicellular tumor spheroids \cite{Freyer} and was previously used in other mathematical models \cite{Kansal,RamisConde}. The migrating cell moves to one of its nearest-neighbor sites chosen at random, interchanging its position with a normal or necrotic cell if there exists any. If the interchanged normal or necrotic cell moved to a site still occupied by other cancer cells, it is eliminated. 
These rules for cancer growth are similar to those used by Scalerandi et al. \cite{Scalerandi} in a deterministic model with collective cell actions controlled by threshold (step) functions of the nutrient concentration per cancer cell.

\noindent 4-\textbf{Become infected} with a probability

\begin{equation}
P_{inf}=1-\exp{\left[ -\left( \frac{\sigma_v}{\sigma_c \theta_{inf}} \right)^2 \right]},
\label{pinf}
\end{equation}
an increasing function of the local viral load per cell, controlled by the parameter $\theta_{inf}$. The model assumes perfect viral selectivity for cancer cells, thereby the infection of normal cells by oncolytic viruses is neglected. The number of viruses $n_v$ that infect a given cell is selected from a Poisson distribution, 

\begin{equation}
P(n_v)= \frac{k^{n_v} e^{-k}}{n_v!},
\end{equation}
where $k$ is the typical viral entry. This Poisson law has been observed in cell cultures as a function of the multiplicity of infection (MOI), defined as the ratio between the total viral load injected in the system and the number of target cells \cite{Ozturk}.
The model assumes that an infected cancer cell does not divide nor migrate because its slaved cellular machinery is focused on virus replication. It is also assumed that infected cancer cells sustain their metabolism until lysis and die only by lysis. The death by lysis occurs with a probability

\begin{equation}
P_{lysis}=1-\exp{\left( -\frac{T_{inf}}{T_l}\right) },
\label{plysis}
\end{equation}
where $T_{inf}$ is the time elapsed since the cell infection and $T_l$ is the characteristic time for cell lysis. The lysis of each infected cancer cell releases

\begin{equation}
v_0=b_s \frac{n_v}{A+n_v}
\end{equation}
free viruses to the extra-cellular medium. Here, the maximum virus burst size $b_s$ and $A$ are model parameters. At the time of lysis, the new free viruses remains on the site of the lysed cell. They diffuse independently through the tissue by performing lattice random walks comprising $q$ steps and are cleared at a rate $\gamma_v$ at each time step. The clearance rate $\gamma_v$ embodies the complex innate and adaptive immune responses to a virus. Such response involves the synthesis of antiviral cytokines, activation/selection of immune cells, and production of antiviral antibodies \cite{Weinberg}.

The tumor starts to grow from a single cancer cell and the therapy begins when the tumor attains $N_0$ cells. It consists of a single direct intratumoral administration in which $N_0 \times MOI$ viruses are uniformly spread over the entire tumor. This approach corresponds to the experimental protocols used in severe combined immune deficient (SCID) mice \cite{Coffey} and in vitro assays \cite{Bischoff,Raj}. Indeed, in most of the in vivo virotherapy experiments the virus are injected directly into a subcutaneous, avascular tumor developed in mice.

More details concerning the procedures used for model simulations are provided in appendices A and B.

\section{RESULTS}
We are interested on the effects of virotherapy on compact, papillary and disconnected tumors, the general morphologies observed in solid malignant neoplasias. Typical patterns corresponding to such morphologies were simulated using the multiscale model for the growth of avascular tumors studied in reference \cite{Ferreira2002}. They are shown in Figure $2$ and the model parameters used are listed in Table 1.

Typical progress curves for cancer cells and free viruses in a simulated virotherapy of a diffuse tumor are shown in Figure $3$. Similar time evolutions are also observed for solid and papillary neoplastic morphologies (see Appendix C). Two regimes are observed: either the cancer cell population keeps growing after the virus administration or both cancer cells and oncolytic viruses are eradicated. In the former regime, in which the virotherapy fails, the viruses can be either eradicated or coexist with the tumor cells. Such coexistence can exhibit oscillations in cancer cells and virus populations due to successive rounds of infection unable to eliminate the tumor. As previously reported \cite{Paiva}, it is worth to emphasize that any of these behaviors can randomly emerge as the response of a given tumor to the virotherapy. Hence, it becomes imperative to determine for each tumor morphology the most probable prognosis after the treatment as well as the chances for the occurrence of the other tumor responses as functions of the virotherapeutic parameters.

The probabilities for the eradication of a papillary tumor are shown in Figure $4$. As one can see, a single intratumoral administration of an aggressive virus ($\theta_{inf}=0.01$) at a $MOI=1.0$ with viral entry $k=1.0$, low clearance rate $\gamma_v=0.03$ and $N_0=10,000$ can eradicate the tumor. Even viruses with low replicative potential, associated to small burst sizes ($b_s=10$, for instance), have almost $100\%$ of efficacy if they spread very slowly ($q\lesssim 4$) throughout the papillary tumor. As shown in Figure $4$(a), the anti-tumor efficacy of low-diffusive viruses depends very weakly on $T_l$, the characteristic time spent by the virus to induce the lysis of infected cancer cells. Furthermore, the anti-tumor efficacy decreases rapidly with the increase of the virus diffusivity, almost vanishing for $q \gtrsim 9$. In contrast, for a highly replicative virus (burst size $b_s=100$), the tumor eradication is almost always certain. The exceptions are for highly cytolytic ($T_l \lesssim 10$) and diffusive ($q \gtrsim 20$) viruses. Indeed, as shown in Figure $4$(b), the probability of therapeutic success goes to zero if the oncolytic virus spreads very fast on the tissue and quickly kills cancer cells.

Bearing in mind the current paradigm (fastest viral replication, cytolysis and spreading), the correlations revealed here among therapeutic outcomes and the traits of the anti-tumor vectors figure out a counter intuitive scenario. Indeed, a low viral diffusivity, independently on the cytolytic period $T_l$ and replicative capacity $b_s$, is sufficient because papillary malignant neoplasias grow slowly. Nonetheless, one can argue that virus exhibiting low diffusion rates in tissues will hardly reach the tumor if intravenously or systemically administered. Hence this trait should be clinically avoided unless delivery barriers have been overcame through some ``pro-diffusive'' strategy. As an alternative, our results indicate that fast spreading viruses combining high replication potentials and slow citolysis can successfully eliminate papillary tumors.

The aforementioned results should be confronted with those obtained for a compact tumor in order to determine if the most effective virus against papillary tumors is equally efficient in the treatment of a compact tumor. Compact morphologies are generated in our model when competition for nutrients is weak and, consequently, the tumor grows fast. The same oncolytic virus and treatment protocol successfully used for papillary tumors ($\gamma_v=0.03$, $\theta_{inf}=0.01$, $MOI=1$, $k=1$ and $N_0=10,000$) fails for compact tumors. They are not eradicated for any $T_l$ and $q$ values studied even for a large burst size ($b_s=100$). So, very efficient viruses against papillary tumors can be ineffective to eradicate compact tumors. While the number of cancer cells increases, the viruses either become extinct, if $q=1$, or coexist with the growing tumor, if $q >1$. In the coexistence regime, the tumor grows continuously but at a lower rate.

Given that solid tumors grow faster than ramified ones, it seems intuitive to assume that more diffusive viruses might be more effective against the neoplastic mass. However, the success probabilities remain null if only the viral diffusivity is increased up to values so large as $q=900$. One alternative towards a successful therapy is reinforce the viral load initially administered. Nonetheless, even at larger $MOI$s ($MOI=5$, for instance) the therapy still completely fails if the viral entry $k=1$ is maintained. The value $k=1$ implies that in average only one virus invades a cell at each infection event.

A remaining strategy to control compact tumor growth is enhance the viral entry. As one can see in figure $5$(a), a large viral entry ($k=5$) has a dramatic effect on the therapeutic success. Indeed, the virotherapy only fails, even using a $MOI=1$, for very low viral diffusivity ($q\le 2$) and large lytic period ($T_l \gtrsim 15$). Furthermore, a significant success is achieved by reducing $k$ while, in balance, raising the $MOI$ used. This is shown in figure $5$(b) in which $k=1.5$ and $MOI=1.5$. In this case, a successful therapy demands an oncolytic virus with appropriated diffusivity and lytic cycle. An adequate diffusion ($3 \lesssim q \lesssim 16$) allows viral spreading throughout the compact tumor at a rate similar or greater than that of the growing neoplasia. But fixed the viral diffusivity, the time $T_l$ must lie within a certain range, as for instance $7 \lesssim T_l \le 32$ for $q=15$. Indeed, a very rapid cytolysis promptly forms ``voids'' containing the majority of the newly released viruses, which impair the generation of new infection waves. In turn, a very long $T_l$ generates successive waves of infection at a low frequency unable to destroy the tumor. These results are consistent with those reported on reference \cite{Paiva}, supporting the robustness of both models.

At last, the response of diffuse tumors to virotherapy was investigated. In our model those disconnected patterns emerge for highly motile cells that uptake nutrients at moderate rates. Considering an oncolytic virotherapy based on a viral agent with high replicative capacity, the probability for the eradication of a diffuse tumor is shown in Figure $6$. As one can see, the success probability also vanishes for highly diffusive viruses and increases significantly for viruses spreading slowly. Optimal oncolytic viruses for treat diffuse tumors should combine high replicative rates and intermediate diffusivity. If the virus has a slower spreading, then a short time for induce the lysis of infected cancer cells is demanded in order to maximize the therapeutic success. At larger viral spreading ($8 \lesssim q \lesssim 16$) the success probability exhibit a re-entrant behavior characterized by an initial decrease, followed by an increase of the therapeutic success as the cytolytic period $T_l$ increases.

\section{DISCUSSION} 
In this work we investigated the oncolytic virotherapeutic outcomes for solid (compact, papillary and diffuse) tumors using numerical simulations. We focussed our study on vectors with high infectivities and small clearance rates corresponding to a severe suppression of the host immune response. Furthermore, since in the present framework the discrete nature of the oncolytic virus allows a more realistic modeling of viral infection and replication processes, the central roles of viral burst size $b_s$, virus entry $k$ and initial MOI for the therapeutic success are highlighted.

Figure $3$ illustrates our typical results: after a single intratumoral virus administration, these tumors can either be completely eradicated or keep growing with time despite a transient remission, as previously obtained for compact tumors \cite{Paiva}. Furthermore, the last behavior can be either a monotonic or an oscillating growth. It is worthy to notice that such oscillations were observed in human myeloma xenografts induced in mice treated with measles virus (MV) \cite{Myers}, in an ovarian cancer xenograft model \cite{Peng}, and in a mathematical approach used by Dingli \textit{et al.}  for modeling MV virotherapy \cite{Dingli}. The therapeutic outcome depends on both viral characteristics and tumor dynamics. So, for fixed viral infectivity $\theta_{inf}$ and persistence within the tissue determined by $\gamma_v$, the key parameters that control treatment success are the virus entry $k$, viral diffusivity $q$, characteristic lytic time $T_l$ and tumor growth rates.

Concerning the virotherapy of solid tumors, two main results deserve special attention. Firstly, in the case of a low virus entry ($k=1$) successful therapeutic outcomes were observed only at large, but realistic, initial MOIs and viral burst sizes. It must be mentioned that the viral burst size used in reference \cite{Paiva}, equal to the initial viral load, was one order of magnitude greater than the larger viral burst sizes observed for real oncolytic viruses. In turn, the large initial MOIs required by successful virotherapies in our present simulations with $k=1$ raises the question of the clinical risks involved. Viral doses ranging from about $10^7$ pfu ($MOI \approx 0.08$) to $10^9$ pfu ($MOI \approx 8$) were used in experimental tests \cite{Ziauddin}. Also, it is known that the oncolytic adenovirus dl1520 (or ONYX-015) is well tolerated at the higher practical administered doses ($2 \times 10^{12}$---$2 \times 10^{13}$ particles) \cite{Relph}. But at very high viral titer (about $1 \times 10^{14}$ particle per Kg) this virus caused the death of a gene therapy patient with ornithine-cystosine transferase deficiency \cite{Lehrman}. Hence, the use of reduced viral doses is pursued in clinical applications, particularly in order to allow for systemic administration.

Successful outcomes could be accomplished by raising the viral entry $k$ (see figure $5$). Indeed, $k$ represents the average number of virus entering in a cancer cell at infection. Thus, large values of $k$ result in a high production of new viruses ($v_0 \sim b_s$, the viral burst size) that are released after cell lysis. This local enhancement of the free virus load enlarges, in turn, the probability of further cancer cell infection. The consequence of such a positive feedback cycle is that the initial viral load is sufficient to trigger a powerful first wave of infection, whose successive rounds ends to destroy the tumor. Moreover, as shown in figure $5$(b), a relatively slight and simultaneous increase of both virus entry and $MOI$ can be enough to guarantee a successful oncolytic treatment of compact tumors with safe viral doses.

Secondly, considering that large viral loads involve risks and that an elevated virus entry may represent a significant technical barrier, treatments based on intermediate $MOI$s and $k$ values become of special interest. In this range ($MOI \sim 1$ and $1\le k<2$), the key factors determining the therapeutic success are viral diffusivity and cytolytic period. The probability of tumor eradication vanishes for viruses that diffuse very fast even at large MOIs. Indeed, a very rapid spreading disperses the viruses outside the growing tumor where they can not trigger new waves of infection. In turn, since the growth of compact tumors is faster, an effective oncolytic virus must not diffuse slower than the tumor spreading. At intermediate values of $k$ and $MOI$, in addition to be able to spread at a rate similar to that of the growing tumor, the virus must induce the lysis of infected cancer cells neither too slowly nor too fast. If the time $T_l$ for cytolysis is very short, ``voids'' containing the majority of the new released viruses are rapidly formed throughout the tumor. These viruses can not trigger new infection waves and cancer cells at the border of some voids escape. In turn, if $T_l$ is very long, the successive waves of infection are generated at a low frequency, excessively distant in time to destroy the tumor. These findings are consistent with the predictions in Figure 5 of reference \cite{Paiva} which suggests that the ideal oncolytic virus should have neither a very short nor a long lytic cycle.

Summarizing, a successful virotherapy of compact tumors demands the simultaneous optimization of several factors: the use of oncolytic viruses with high replicative capacities (burst sizes), adequate diffusivities (rapid but not too rapid spreading) and lytic cycle (neither too short nor too long), as well as a  large initial viral entry. The importance of the initial MOI, strongly emphasized by Demers et al. \cite{Demers} and Myers et al. \cite{Myers}, is much smaller than that associated to the virus entry. In those experiments, huge variations in the initial viral load, from $10^4$ to $10^{10}$ viral particles, were performed. Such values correspond to $MOI$s ranging from around $0.5$ to about $500$. As one can intuitively expect, our simulations evidence that a huge $MOI$ eradicate the tumor independently of other factors (viral traits and tumor growth dynamics). However, for virotherapies based on low virus entries and $MOI$s there is, fixed the viral diffusivity, an optimal range for its lytic cycle period because the virotherapy fails if the cytolysis is either too short or too long. This result is in complete agreement with the findings of Paiva \textit{et al.} \cite{Paiva}. Furthermore, we point out that the virotherapeutic efficacy is much more sensitive to other parameters, for instance viral entry and burst size, than to $MOI$.

With regard to papillary tumors, a virotherapy based on small virus entry ($k=1$) and initial viral load ($MOI=1$) can control cancer growth. In this case the key factors determining a successful therapeutic outcome are the virus spreading and its replicative ability. Indeed, for small viral burst sizes the therapeutic outcome for papillary tumors is almost independent on $T_l$. We find that the virotherapy fails for oncolytic viruses with high diffusivity. Success probabilities greater than $70\%$ are observed only for very small diffusivities ($q \lesssim 6$). In contrast, for a fixed diffusivity, the therapeutic success of highly replicative viruses exhibits a re-entrant behavior as a function of their lytic periods, as shown in Figure $4$(b). Again, our simulations suggest an optimal range for the time $T_l$ demanded by the oncolytic virus to kill an infected tumor cell.

At last, for fixed small virus entry ($k=1$) and initial viral load ($MOI=1$), optimal oncolytic viruses for treat diffuse tumors should exhibit high replicative rates and intermediate diffusivity. For faster viral spreading ($q \gtrsim 8$), a successful therapeutic outcome has a complex dependence on $T_l$. A re-entrant behavior is observed, leading to a decreasing success probability at an intermediate range of the time $T_l$ for cytolysis. In contrast, for oncolytic viruses having very slower spreadings, highly successful therapies demanded short times for induce the lysis of infected cancer cells.

A relevant issue is how our simulation results can be compared with those obtained from experimental assays or clinical trials on oncolytic virotherapy. Unfortunately, this represents a difficult task due to the scarcity of quantitative experimental data. Most of the studies on oncolytic virotherapy focused primarily on the safety of these vectors and evaluate the anti-cancer efficacy of viruses. Tests \textit{in vitro} involves cancer cells cultured in monolayer or 3d compact tumor spheroid models. Either the xenograft models in experimental animals or the human tumors treated in clinical trials are also solid tumors with compact morphologies. Furthermore, since very distinct tumors were treated using several different viruses, both the system and the therapeutic conditions varied widely. Hence, almost all parameters have changed simultaneously from one experiment to another. Basically, the results are qualitative: the anti-tumor efficacy of viruses has been limited. Virotherapy is impaired by multiple factors, specially the low viral loads and its suboptimal delivery to the target site, barriers to viruses spreading within the tumor and the host immune response against these vectors.

Our simulations reveal that the virus entry $k$ is the major parameter determining the therapeutic outcome. Once its value is fixed, the lower is the initial dose of virus administered, smaller is the chance of tumor eradication. Experimental \cite{Myers} and mathematical \cite{Dingli} results support this direct correlation between MOI and therapeutic success. Also, our simulations indicate that virus replication within the tumor tissue, modelled by the viral burst size $b_s$, is the next main trait in the hierarchy of factors determining the virotherapeutic success. Such finding is supported by observations from Friedman \textit{et al.} \cite{Friedman}. Decreasing in the hierarchy, the viral diffusivity emerges as the next key parameter for a successful virotherapy. Experimental observations reporting that antitumor efficacy increases as intratumoral virus spreading is enhanced due to voids and channels generated by induced cancer cell apoptosis inside a tumor \cite{Nagano} are in agreement with this result. Finally, experimental data that confront our results concerning the effects of the cytolytic time on the therapeutic outcome are lacking.

From our model simulations the efficiency of virus entry in a cell, its replicative capacity inside an infected cell, and its diffusivity within the tumor tissue emerges as three crucial factors to fuel and amplify the successive rounds of infection necessary to control or eradicate the tumor. Hence, it is worthy to briefly comment about some trends on the experimental research in these directions. Concerning virus entry, a major challenge is the engineering of entry proteins to achieve higher infectivity. But this approach is diametrically opposite to the search for entry inhibitors, neutralizing antibodies and vaccine immunogens that elicit antibodies against virus entry proteins \cite{Dimitrov,Eckert}. Instead of fight virus, the goal is transform it into a better entry machine while reinforcing its retargeting specificity. Here, experimental results are scarce and constrained to comparative analysis of the infectivity of distinct oncolytic viruses. Thus, for instance, Ketola et al. \cite{Ketola} compared the oncolysis in five human osteosarcoma cell cultures infected with the Semliki Forest virus $VA7-EGFP$ and the adenovirus $Ad5\Delta 24$. They found that the kinetics of adenovirus infection was much slower than that of Semliki Forest virus. High amounts of adenoviruses were required to assure their spread throughout the cell culture and to lyse all the osteosarcoma cells. Also, the authors demonstrated differential efficacy of the Semliki Forest virus against the human osteosarcoma cell lines. Virus spreading was observed in four of five cell lines studied. In particular, the cell lineage $MG-63$ seemed to be relatively resistant to the $VA7-EGFP$ infection. Clearly, these results can be understood in terms of differences in virus entry and replicative capacity as evidenced in our model. By the way, some attempts to enhance the replication of oncolytic viruses have been performed since, for instance, the burst size of a wild-type HSV is one or two orders of magnitude greater than that of the engineered vector hrR3. Alternatives include to place viral genes under the transcriptional control of tumor specific promoters \cite{Chung} or mutate viruses to express genes that overcome tumor cell pathways blocking viral protein synthesis \cite{Mulvey}. Again, more attention need to be paid to this approach. A greater activity is observed in the field of virus spreading. Several vasoactive cytokines, or physical treatments, such as radiation or heat, increase tumor vascular permeability and blood flow, leading to faster diffusion \cite{Jain1990}. Such strategies can be particularly relevant for systemic virus administration and to treat tumors at the vascular stage, mainly compact tumors exhibiting a fast growth as our results indicate. The incorporation of lytic enzymes, such as hyaluronidase, into a conditionally replicative adenovirus could also increase diffusion rates \cite{Alemany}. Furthermore, agents that decrease the glycosaminoglycan concentration \cite{Mok2007} or that degrade interstitial collagen \cite{Netti,Brown} can improve transport in tumors. For instance, tumor collagen can significantly hinder diffusion, and it was shown \cite{Mok2007} that the matrix metalloproteinases$-1$ and $-8$ can modulate the tumor matrix and improve the distribution of an oncolytic virus throughout the tumor, without affect cell proliferation or viral replication. In addition, a previous work \cite{Montel} showed that the matrix metalloproteinase$-8$ is antimetastatic, so it could be safely used to improve the viral diffusion in tumors. Strategies that change the tumor microenvironment are specially important to enhance the efficacy of virotherapy against avascular tumors and occult dormant metastasis.

In summary, the outcomes of oncolytic virotherapy on solid tumors were investigated through computer simulations of a multiscale model for cancer growth. The model combines macroscopic diffusion equations for the nutrients and stochastic rules for the actions of individual cells and viruses. Our simulations reveal that in immunosuppressed hosts the antitumor efficacy of a virus is determined primarily by its efficiency of entry, its replicative capacity within the cancer cell, and its ability to spread over the tissue. However, the balance between these viral traits depends on the tumor morphology. Indeed, the virotherapy of papillary tumors based on oncolytic viruses with small diffusivities is highly effective (success probability $ \ge 80 \%$) even for small virus entry but high replication capacity. Furthermore, the therapy fails if the virus both spreads and lysis infected cells very rapidly. In contrast, for compact tumors the therapy fails if vectors characterized by small virus entry and slow intratumoral spreading are used at safety $MOI$s. In turn, for diffuse tumors an intermediate viral diffusivity maximizes the therapeutic success. Thereby, the design of efficient oncolytic viruses must take into account the dynamics of tumor growth, which is fast for compact but slow for papillary cancers. It is the tumor growth that sets up the optimal traits for oncolytic viruses. Those traits do not necessarily include high diffusivity and cytolysis, which are naively assumed as necessary conditions to amplify the therapeutic inoculum \textit{in situ} and promote a fast viral spreading throughout the affected tissue.

\begin{acknowledgments}
This work is dedicated to the memory of our friend and collaborator in this enterprise professor Marcelo Jos\'e Vilela, recently dead by cancer. 
The authors would like to thanks the Brazilian supporting agencies CNPq, Capes and Fapemig. 
\end{acknowledgments}



%


\begin{figure}
\caption{ (Color online) {\bf Schematic illustration of the mathematical approach.} The model consider normal cells, tumor cells, and oncolytic viruses. Uninfected cancer cells can replicate by mitosis or die due to starvation (lack of nutrients) or move or become infected, a quiescent state in which the cellular machinary is slaved to promote viruses replication. Cell actions are stochastically governed by probabilities $P_{div}$, $P_{del}$, $P_{mov}$ and $P_{inf}$ dependent on the nutrient and virus concentrations per cancer cell. Infected cancer cells die by lysis with probability $P_{lysis}$. After mitosis, the new tumor cell can either pile up (loss of contact inhibition) at the site of its parent cell or invade randomly one of their normal or necrotic nearest neighbor cell, if there exists any. Nutrients are modelled as continuous concentrations which evolve in space (tissue) and time according reaction-diffusion equations. Essentially, nutrients diffuse from the capillary vessel through the tissue and are uptaken by normal and tumor cells at distinct rates. Free viruses perform Brownian motions (lattice diffusion) and are cleared at a rate $\gamma_v$. The source of new viruses are infected cancer cells which release $v_0$ new viruses after undergoing lysis. The virus clearance rate represents the simplest way to introduce viral loss due mainly to the innate and adaptive immune response mediated by antibodies, CD8 T cells, interferons and other cytokines or inherent viral instability.}
\label{model_scheme}
\end{figure}

\begin{figure}
\caption{ {\bf Simulated morphologies for avascular tumor growth.} (A) Solid or compact tumors grow exponentially fast under weak competition for nutrients and reduced cell motility. (B) Diffuse or disconnected tumors are characterized by a moderate level of nutrient competition and a high cell motility. (C) Papillary patterns emerge under strong nutrient competition and low cell motility, resulting in a very slow growth. The parameters used are listed in Table 1. All tumor patterns shown have $10^4$ cells. The capillary is at the bottom of the frames.}
\label{patterns}
\end{figure}

\begin{figure}
\caption{(Color online) {\bf Temporal evolution of uninfected cancer cells (solid black line) and free viruses populations (dashed blue line) in a diffuse tumor.} A single intratumoral virus load was administered when the tumor had $N_0=5,000$ cells. Notice that three different behaviors were obtained using exactly the same fixed set of parameters and initial conditions ($MOI=1$, $bs=100$, $k=1$, $q=25$, $T_l=16$). These different behaviors emerge from stochasticity. The parameters used are listed in Table 1. }
\label{cancer_pops}
\end{figure}

\begin{figure}
\caption{ {\bf Papillary tumor erradication probability for distinct virus diffusivities $q$ and time for lysis $T_l$.} Two viral burst sizes (A) $bs=10$ and (B) $bs=100$ are shown. The values $\gamma_v=0.03$, $\theta_{inf}=0.01$, $MOI=k=1$, and $N_0=10^4$ are fixed. The probabilities were evaluated from $50$ independent samples.} 
\label{success_probR}
\end{figure}

\begin{figure}
\caption{ {\bf Solid tumor erradication probability as a function of the parameters $q$ and $T_l$.} (a) $MOI=1$ and viral entry $k=5$, and (b) $MOI=k=1.5$. The viral burst size is $b_s=100$ and the values $\gamma_v=0.03$, $\theta_{inf}=0.01$, and $N_0=5,000$ are fixed. The probabilities were evaluated from $50$ independent samples for each pair ($q$, $T_l$).}
\label{success_S_moi15}
\end{figure}

\begin{figure}
\caption{ {\bf Diffuse tumor erradication probability as a function of the parameters $q$ and $T_l$.} A viral burst size $bs=100$ was used. The values $\gamma_v=0.03$, $\theta_{inf}=0.01$, $MOI=1$, $k=1$ and $N_0=10,000$ are fixed.}
\label{success_probD}
\end{figure}

\newpage

%


\appendix

\section{Simulation protocol}

The simulations were implemented as follows. At each time step, eq. ($1$) is numerically solved in the stationary state ($\partial \phi/\partial t = 0$) through relaxation methods. Also, a fraction of the free viruses present in every site is cleared, and each remaining virus performs an independent random walk with $q$ steps. Provided the nutrient concentration and viral load at any lattice site, $N_c(t)$ cancer cells are sequentially selected at random with equal probability. (Here, $N_c(t)$ is the total number of tumor cells, uninfected or infected, at the time $t$.) For each one of them, a tentative action (division, death, migration or infection for an uninfected cancer cell, and lysis for an infected one) is randomly chosen with equal probability. The selected cell action will be implemented or not according to the corresponding local probabilities determined by eqs. ($2$), ($3$), ($4$), ($5$) and ($7$), and the time is incremented by $\Delta t = 1/N_c(t)$. If the carried action is the infection of the selected cell, an integer random number $n_v$ distributed as a Poissonian is generated and compared with the local virus population $\sigma_v$. If $\sigma_v \ge n_v$, $n_v$ viruses invade the uninfected cancer cell, decreasing $\sigma_v$ by $n_v$. Otherwise, this process will be repeated until generates a $n_v \le \sigma_v$. In turn, if the carried action is the lysis of the selected cell, $v_0$ new viruses are introduced at the site of the lysed cell. At the end of this sequence of $N_c(t)$ updates, a new time step starts and the entire procedure (solution of the reaction-diffusion equations, virus clearance and spreading, and cell dynamics) is iterated. It is worthy to notice that a particular cell can possibly perform more than one action in a certain time step, but in average every cancer cell will perform the same number of actions. Also, the increment time $\Delta t$ assures that, in average, each cancer cell present in a given time has a chance to perform an action during that time step. The simulations stop if any tumor cell reaches the capillary vessel or the tissue border or if the tumor is erradicated ($N_c=0$).

\section{Parameter estimates}

In all simulations, the tissue is represented by a square lattice of linear size $L=500$ and lattice constant $\Delta=10$ $\mu$m, corresponding to a tissue section of about $25 mm^2$. Assuming a DNA synthesis phase of the cell cycle of about $11$ h \cite{Rew}, one time step in the simulations corresponds to about $4-5$ h. The parameters $\theta_{div} = 0.3$ and $\lambda_2= 10$ were fixed. In order to generate compact, papillary, and disconnected tumor growth patterns distinct values of the parameters $\theta_{del}$, $\theta_{mov}$, $\lambda_1$, and $\alpha$ were used, as listed in Table 1. The parameters $\gamma_v$ and $\theta_{inf}$, characterizing the oncolytic virus, were also fixed (Table 1). The key parameters controlling virus diffusion $q$, entry $k$, replication $b_s$, its time for triggering cell lysis $T_l$ and $MOI$ specifying the virotherapy, were varied. The therapy begins when the tumors have either $N_0 = 5000$ or $10,000$ cells.

The value fixed for $\gamma_v$, controlling viral clearance, is the same used in Ref. \cite{Paiva} and corresponds to a virus with a long residence time in the tissue (low clearance rate). In turn, the value $\theta_{inf}=0.01$ corresponds to a highly infective oncolytic virus ($P_{inf}=90\%$). Both of than are rather arbitrary, but for comparision, $70\%$ of the $hrR3$ viruses successfully invade uninfected glioma cells \cite{Friedman}. The other viral parameters were varied in ranges biologically reasonable. Indeed, $b_s$ ranged in the interval $[10,100]$ as experimentally observed for the burst size of the oncolytic virus $hrR3$ \cite{Friedman}. The MOI also varied within ranges used in virotherapy trials \cite{Ziauddin}. For comparision, viral doses ranging from about $10^7$ pfu (MOI $\sim0.08$) to $10^9$ pfu (MOI $\sim8$) were used in experimental tests \cite{Ziauddin}. The characteristic time for lysis $T_l$ was varied as in ref. \cite{Paiva}. Assuming a simulational time step of $4$ hours, $T_l$ ranged from $8$ hours to $5$ days. For comparision, the mean life time of glioma cells infected by $hrR3$ is $\sim18$ hours \cite{Friedman} and the lytic cycle of the Adenovirus $Ad2$ is $32-36$ hours. So, the simulations assume viruses whichare $4$ times faster to $4$ times slower than the $Ad2$. The values considered for the virus entry $k$ lead, in average, to cell infections by few virus simultaneously. Unfortunatelly, we can not found direct measures or estimates fo $k$ in the literature. Finally, $q$ (the number of steps in a viral random walk), a measure of the viral spreading capacity, was estimated as follows. Since in a random walk $<r^2>=Dt$ or $<r^2>=q\Delta^2$, where $D$ is the diffusion constant, $t$ is the diffusion time and $\Delta$ the lenght step, the virus diffusivity can be determined from q. Assuming $\Delta \sim10 \mu m$ (lattice constant=linear cell size) and $t=4h$, one finds $D_{min}=2.5 \times 10^{-5}mm^2h^{-1}$ for $q=1$ and $D_{max}=6.25 \times 10^{-4}mm^2h^{-1}$for $q=25$. These are the limit values used for the virus diffusivity in the simulations. Taking as a reference value $D=1.8 \times 10^{-4}mm^2h^{-1}$, the diffusivity of the HSV in tumors with high collagen content \cite{Mok2009}, $D_{min}$ is $\sim14$ times smaller and $D_{max}$ is $\sim3.5$ times greater than this reference.

\begin{table}[!hp]
\caption{\label{tab_par}Parameter values used in the simulations.}
\begin{footnotesize}
\begin{tabular}{cccl} \hline \hline 
Tumor       \hspace*{-0.3cm} & Parameter       & Values         & Description                                          \\
morfology    &                 &                &                                                      \\ \hline
             &  $\theta_{div}$ & $0.3$ (fixed)  & Controls the probability    \\
             &				   &				& of cancer cell division.\\
             &  $\lambda_2$    & $10$ (fixed)   & Controls the competition            \\ 
                          &				   &	&	for NE nutrients.		 \\ \hline

          & $\theta_{del}$  & $0.03$         & Controls the probability       \\
              &				   &			& of cancer cell death. \\
   Compact     & $\theta_{mov}$  & $5$            & Controls the probability    \\
   or              &				   &				& of cancer cell migration. \\
 solid            & $\lambda_1$     & $25$           & Controls the competition             \\
              &				   &				&for E nutrients.            \\
             & $\alpha$        & $0.02$          & Dimensionless nutrient   \\ 
           &				   &				&  uptake rate for normal cells. \\ \hline
   
        Papillary    & $\theta_{del}$  & $0.01$         &                                                      \\
   or & $\theta_{mov}$  & $5$            &                                                      \\
      ramified        & $\lambda_1$     & $200$          &                                                      \\
             & $\alpha$        & $0.06$          &                                                      \\ \hline
 
       Disconnected   & $\theta_{del}$  & $0.01$         &                                                      \\
  or     & $\theta_{mov}$  & $0.001$        &                                                      \\
      diffuse       & $\lambda_1$     & $50$           &                                                      \\
             & $\alpha$        & $0.06$          &                                                      \\ \hline \hline
Viral traits &                 &                &                                                      \\ \hline
             &  $\theta_{inf}$ & $0.01$ (fixed) & Determines the chance of \\
             &					&				& virus infection (infectivity).\\
             &  $\gamma_v$     & $0.01$ (fixed) & Virus clearance rate \\
             &&& on the tissue.                  \\
             &  $A$            & $0.5$ (fixed)  & Controls the number of virus  \\
             &					&				& released after cell lysis \\
             &  $MOI$          & $1,1.5,5$      & Initial viral load \\
             &					&				&(MOI) administered.               \\
             &  $k$            & $1,1.5,5$      & Number of virus entering \\
             	& & &a cell at infection          \\
             &  $bs$           & $10,50,100$    & Maximum number of viruses \\
             &&& released after cell lysis. \\
             &  $q$            & $1,4,9,16,25$  & Number of random steps per-\\
             &&& formed by a virus per MCS. \\
             &  $T_l$          & $2,4,8,16,32$  & Characteristic time \\
             &&& for cell lysis.                  \\ \hline \hline 
\end{tabular}
\begin{flushleft} Abbreviations: NE and E, non-essential and essential for DNA replication and mitosis, respectively; MOI, multiplicity of infection; MCS, Monte Carlo or simulation time step.
\end{flushleft}
\end{footnotesize}
\end{table}

\section{}

Here, additional results concerning the progress curves of virus and uninfected cancer cells for compact and papillary tumors are shown in figures $7$ and $8$, respectively. Also, the effect of the viral burst size for oncolytic agents characterized by moderate ($k=2$) and high ($k=5$) virus entry are exhibited in figures $9$ and $10$, respectively.

\begin{figure}[!ht]
\caption{ (Color online) {\bf Progress curves - $N_c$ (solid black line) and $v_{free}$ (dashed blue line) - for compact tumors.} The values $q=4$, $T_l=16$, $b_s=100$ and $N_0=5,000$ were used in all figures. In the case of tumor eradication, $MOI=5$ and $k=5$ were used. Coexistence of tumor cells and viruses is observed at $MOI=1$ and $k=1$. At last, virus erradication occurs at $MOI=0.01$ and $k=0.01$.}
\end{figure}

\begin{figure}[!ht]
\caption{ (Color online) {\bf Progress curves - $N_c$ (solid black line) and $v_{free}$ (dashed blue line) - for papillary tumors.} The values $q=25$, $T_l=4$, $b_s=100$ and $k=1$ for the virus and $MOI=1$ and $N_0=5,000$ for the virotherapy were used.}
\end{figure}

\begin{figure}[!ht]
\caption{ {\bf The success probability of a virotherapy applied to compact or solid tumors using $MOI=k=2$.} At the left $bs=10$ and at the right $bs=100$ were used. $N_0=5,000$ was fixed for the virotherapy. As can be noticed, an unexpected coupling between $T_l$ and $bs$ is observed. Indeed, for a large viral burst size $bs=100$ the success probability is practically independent on $T_l$. However, for a small value ($bs=10$) this probability decreases with increasing $T_l$ for $q<10$ (small viral spreading) but increases with $T_l$ at larger $q$ values. It seems that at a high diffusivity even a virus with a low replicative capacity can sustain successive waves of infection by lysing the scattered cells that survives the previous round of infection.}
\end{figure}

\begin{figure}[!ht]
\caption{ {\bf The success probability of a virotherapy applied to solid tumors using $MOI=1$, $k=5$ and $N_0=5,000$.} At the left $bs=10$ and at the right $bs=100$. Notice that the therapeutic success decreases only for virus with very low spreading, particularly when its replicative capacity is low too.}
\end{figure}

\newpage


\providecommand{\noopsort}[1]{}\providecommand{\singleletter}[1]{#1}%
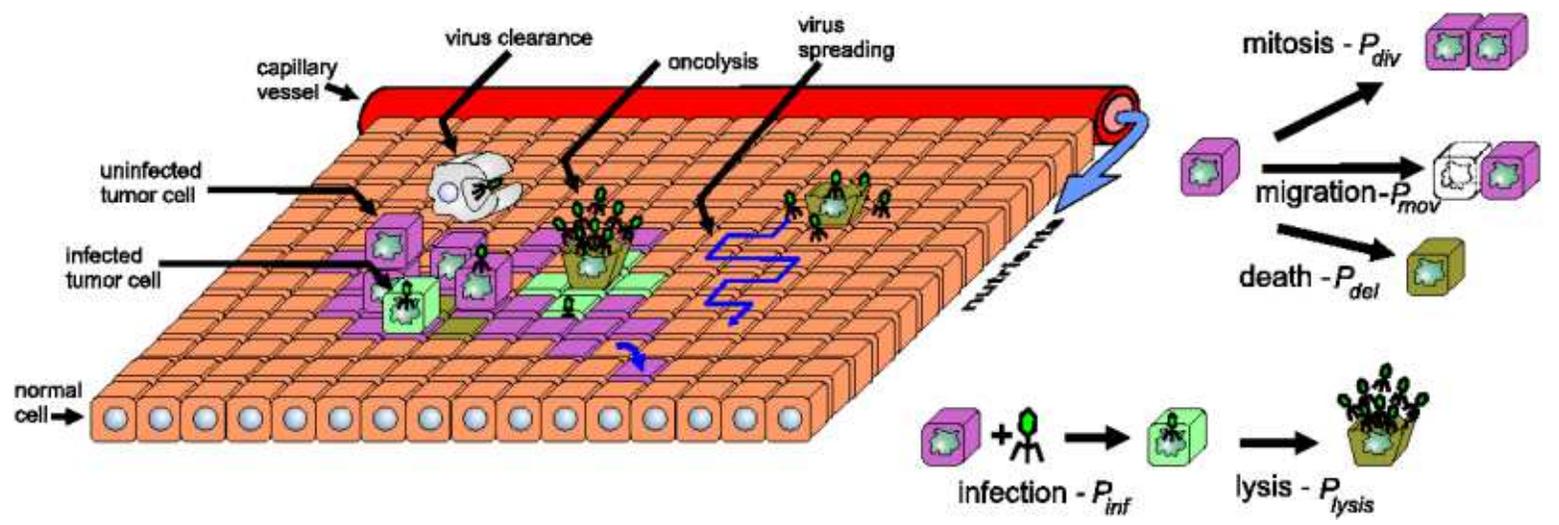
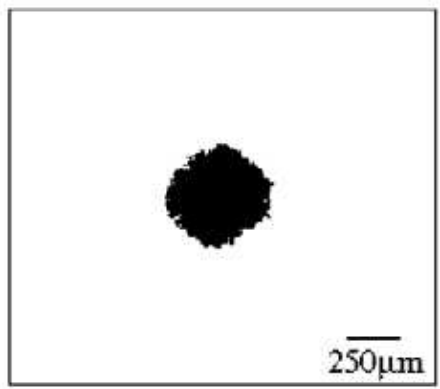
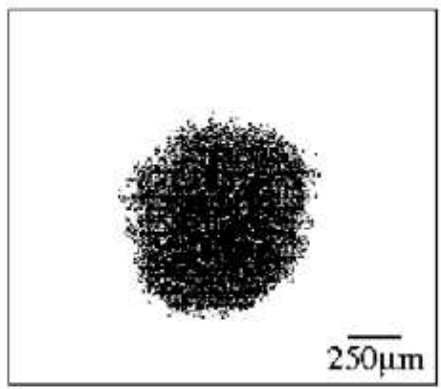
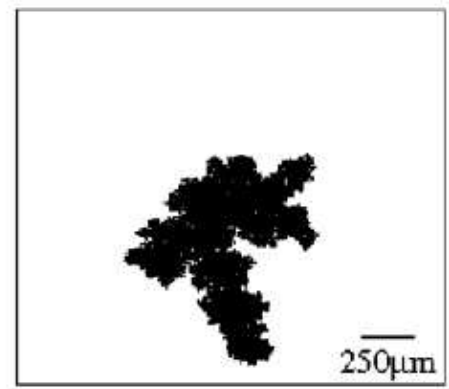
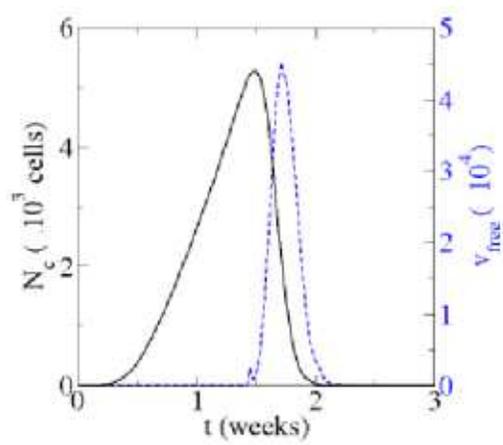
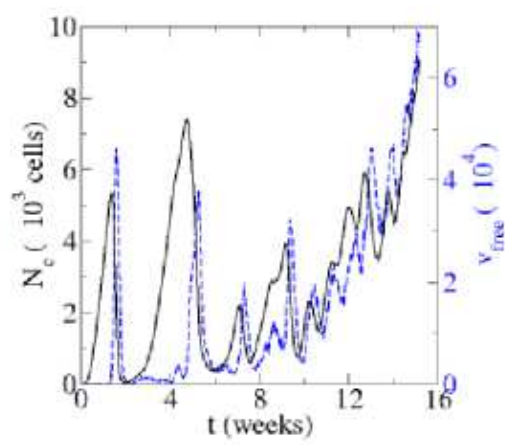
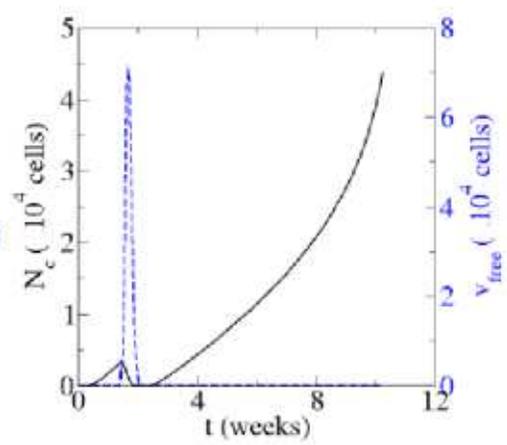
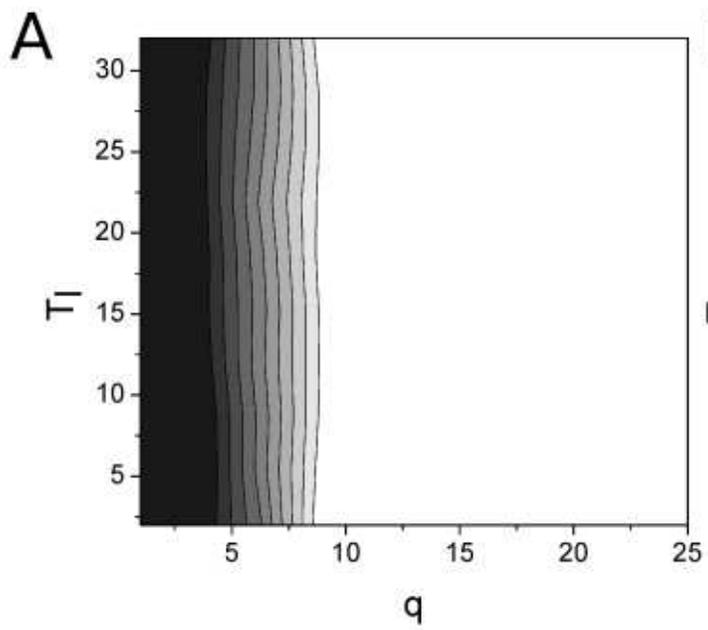
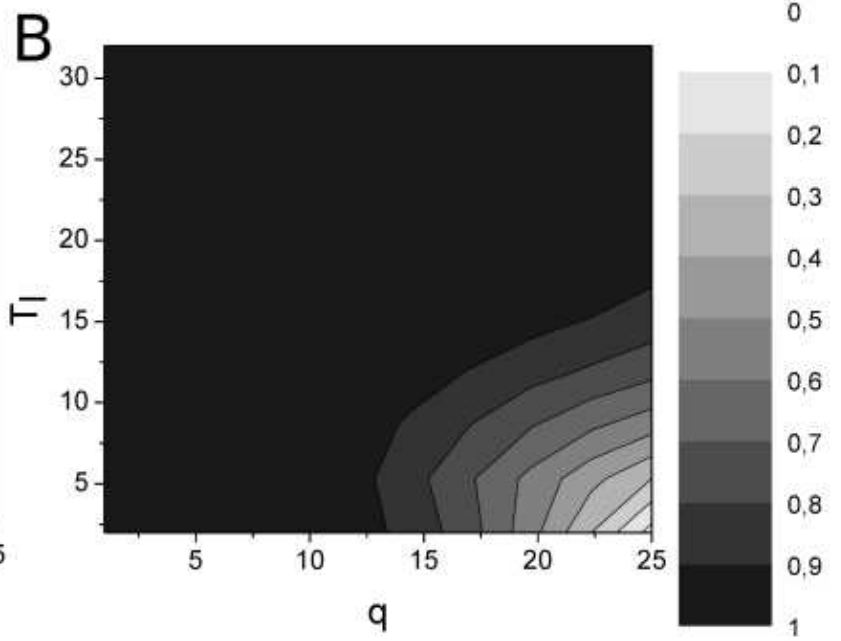
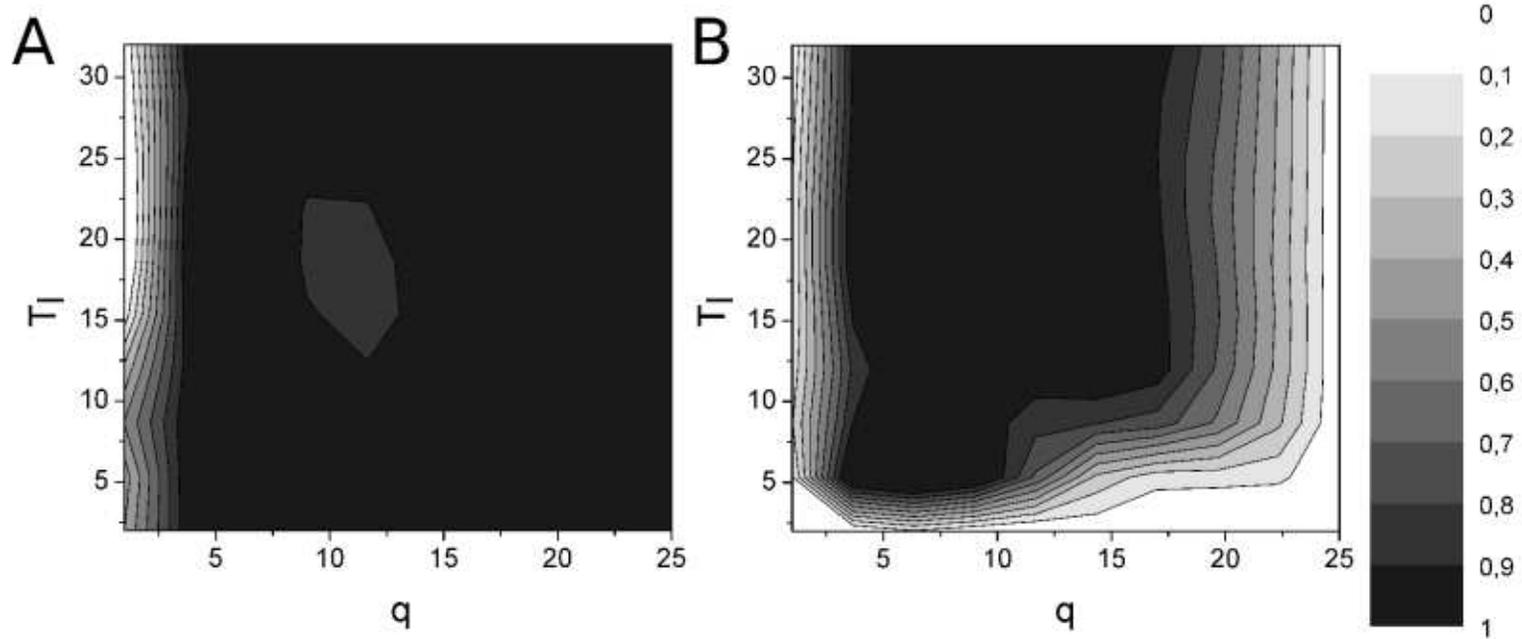
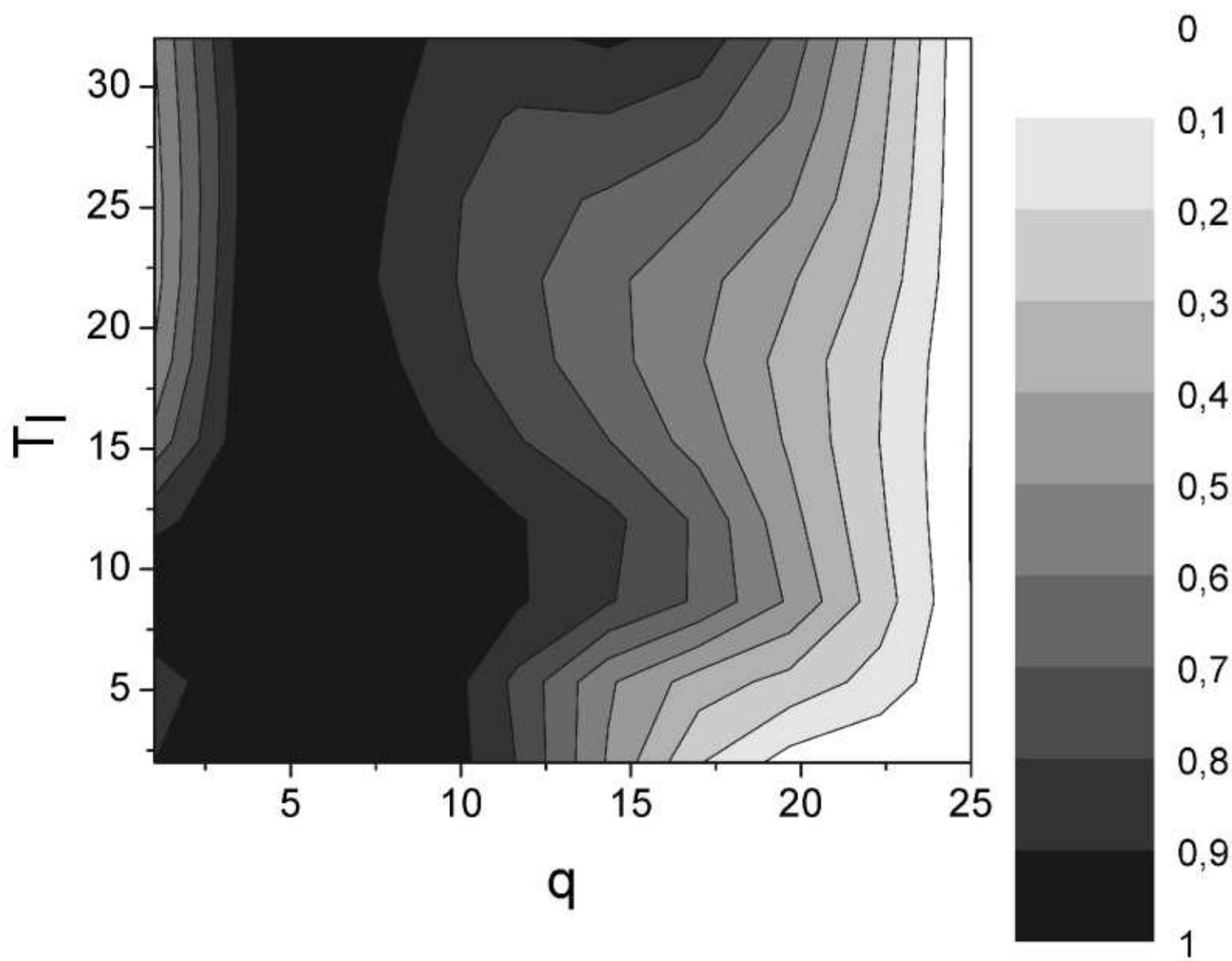
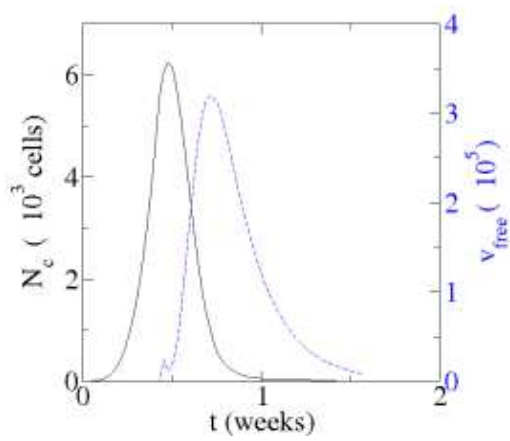
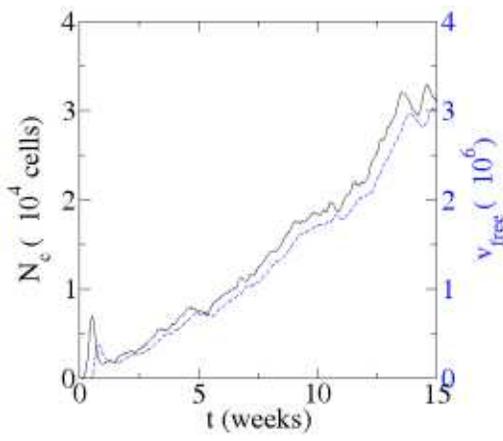
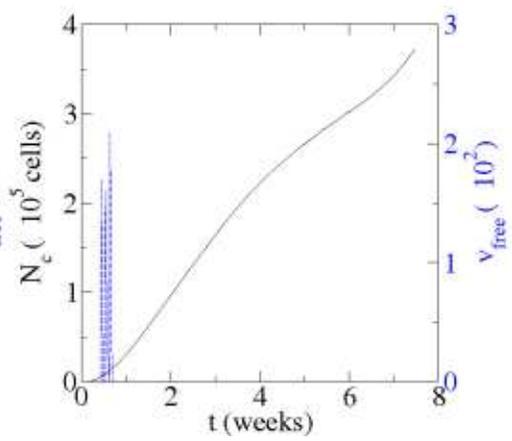
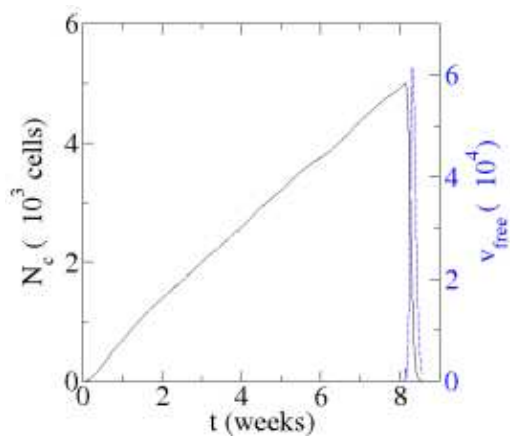
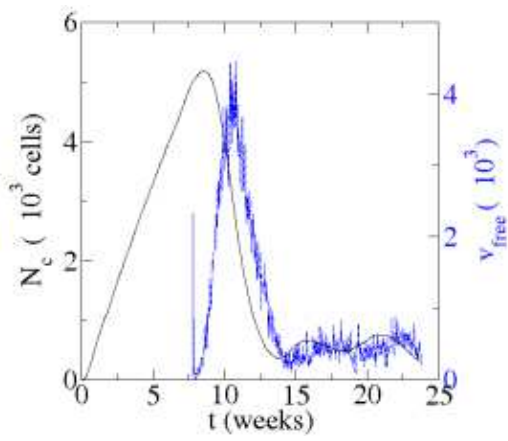
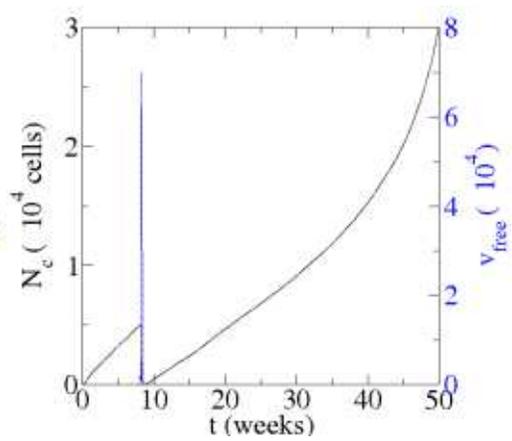
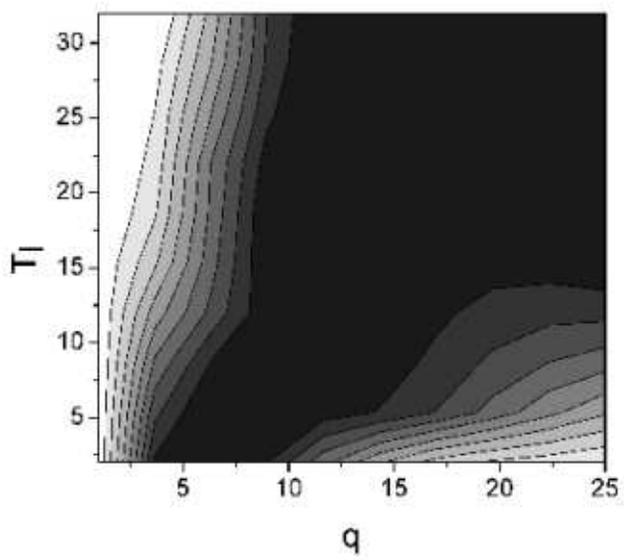
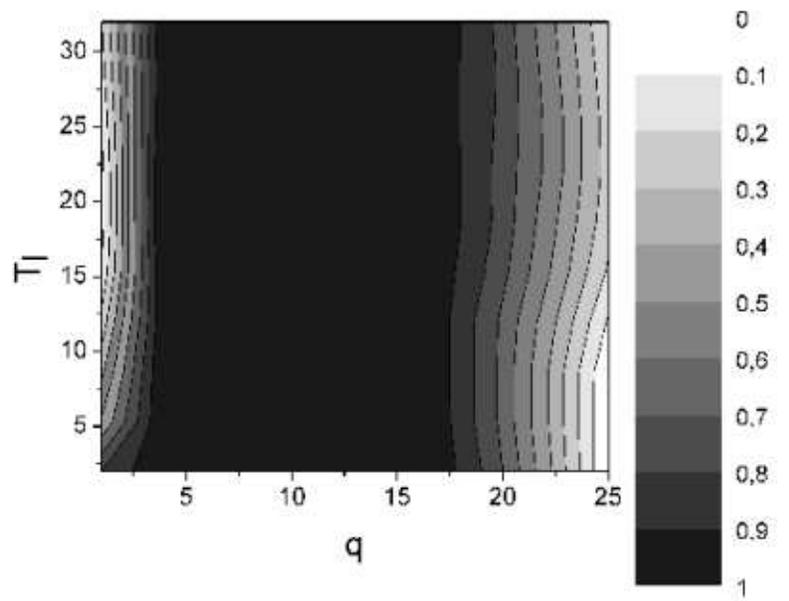
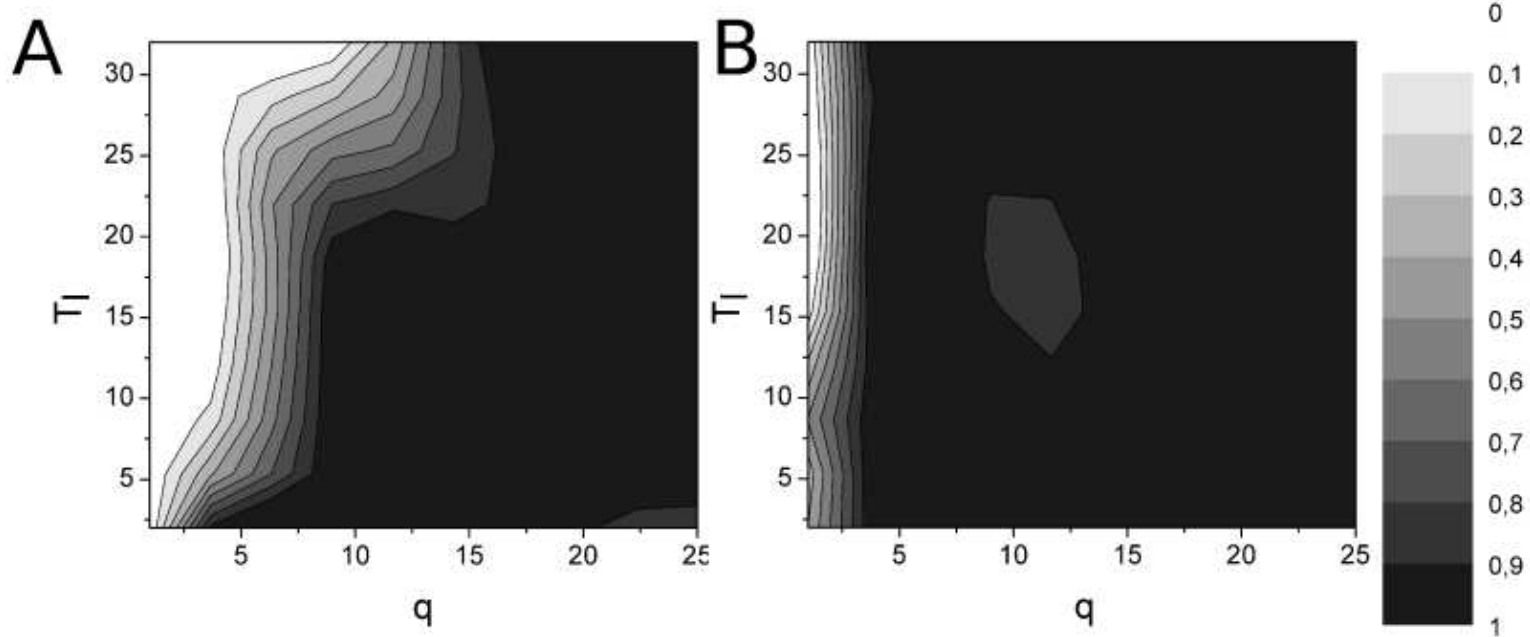
\end{document}